# Comparing Toxicity Across Social Media Platforms for COVID-19 Discourse

Nahiyan Bin Noor, Niloofar Yousefi, Billy Spann, Nitin Agarwal
COSMOS Research Center
University of Arkansas at Little Rock
Little Rock, AR 72204, USA
e-mail: nbnoor@ualr.edu, nyousefi@ualr.edu, bxspann@ualr.edu, nxagarwal@ualr.edu

*Abstract*— The emergence of toxic information on social networking sites, such as Twitter, Parler, and Reddit, has become a growing concern. Consequently, this study aims to assess the level of toxicity in COVID-19 discussions on Twitter, Parler, and Reddit. Using data analysis from January 1 through December 31, 2020, we examine the development of toxicity over time and compare the findings across the three platforms. The results indicate that Parler had lower toxicity levels than both Twitter and Reddit in discussions related to COVID-19. In contrast, Reddit showed the highest levels of toxicity, largely due to various anti-vaccine forums that spread misinformation about COVID-19 vaccines. Notably, our analysis of COVID-19 vaccination conversations on Twitter also revealed a significant presence of conspiracy theories among individuals with highly toxic attitudes. Our computational approach provides decision-makers with useful information about reducing the spread of toxicity within online communities. The study's findings highlight the importance of taking action to encourage more uplifting and productive online discourse across all platforms.

*Keywords-Toxicity analysis; social network analysis; COVID-19; Parler; Twitter; Reddit.*

## I. INTRODUCTION

The most widely used social media platforms, such as Facebook, Twitter, and YouTube, have established community guidelines and enforcement mechanisms to regulate harmful content and misinformation, but free-speech platforms like Parler have been more accommodating towards hate speech, conspiracy theories, and potentially harmful misinformation. Reddit is another social media platform that is basically discussion based; it is a free-speech platform like Parler. However, after the increase of misinformation and hate speech, the policymakers imposed several guidelines and banned some subreddits that spread misinformation, toxicity, and hate speech. Parler is a micro-blogging platform comparable to Twitter that, by design, lacks the content moderation rules and capabilities of the platform it emulates. Parler was created before the emergence of COVID-19, but it has since become an important vector for online misinformation, a place where users can spread COVID-19 misinformation without restrictions. Even though there are multiple guidelines and regulations on Twitter and Reddit to stop people from posting a toxic posts, hate speech, or misinformation, it is not possible to remove toxicity from these platforms.

Managing social media platforms' security is difficult, but examining this harmful content can assist in solving the problem. Our study adds to the current body of knowledge on social media safety.

This paper considers misinformation a claim that contradicts or distorts the common understanding of verifiable facts [1]. Formerly obscure, in 2020, Parler enjoyed a surge in popularity following a push by conservative pundits and politicians to move away from larger, more mainstream social media platforms due to the perception of bias and censorship against conservative viewpoints on those platforms. In 2020, as the COVID-19 pandemic spread worldwide, users of Twitter and the primarily far-right user base of Parler engaged in discussions. They posted content about the vaccination efforts to stop the spread of COVID-19. This work is a comparative analysis of the toxicity of COVID-19-related content on Twitter, Parler, and Reddit from January 1, 2020, through December 31, 2020. Within our text corpus of users' posts, we compared the evolution of the toxicity level over the time frame of analysis. We presented evidence that Reddit contained a higher level of toxicity regarding the COVID-19 discourse than did Twitter and Parler over the four COVID-19-related content datasets we analyzed. From Reddit, among the four COVID-19-related content, the vaccination-related contents are more toxic than any other topic, which makes Reddit the most toxic platform.

This work answers four research questions:
1) How do Twitter, Parler, and Reddit differ about the existence of toxicity within user-generated text content?
2) Of the three platforms, which one contains the highest overall level of toxicity?
3) How did the average toxicity level change over time within Twitter, Parler, and Reddit datasets?
4) Which COVID-19-related topic is the most toxic in which social media?

The remainder of this paper is organized as follows. In Section 2, the related work that has been published regarding toxicity on social media is presented. Section 3 describes the data collection process and the methodology used in this paper. Section 4 presents the highlights from our results and analysis. Finally, Section 5 concludes with the contributions of this work and presents our plans and ideas for future work.

## II. RELATED WORKS

This section will briefly overview toxicity and its spread in social media. Currently, a massive volume of content in social media demands tools and methods to detect toxicity. It will help to prevent the spread of toxicity in social media. Some researchers focused on this domain, and some studies developed a new method for this aim.

Sahana et al. [2] proposed a binary classification for detecting toxic contents; the authors classify toxic comments from non-toxic comments regardless of the nature of the toxicity. A similar approach has been made by Taleb et al. [3] in their research studied of different approaches to detect toxic comments on social media. For this purpose, the authors perform a binary classification to indicate whether a comment is toxic. On the other hand, Kumar et al. [4] suggested classifying toxic comments into various categories; for this assignment, he performs multiple machine learning approaches such as Logistic Regression, K Nearest Neighbors., Bernoulli Naïve Bayes (NB), multinomial NB, Support Vector Machine (SVM), and Random Forest. Comparing these algorithms helps us identify which method performs better in detecting multiple toxicities. Watanabe et al., [5] detected toxicity and hate speech on Twitter, proposed an ML technique using sentiment and semantic-based features.

Gröndahl et al. [6] claimed that current hate-speech detection models are inaccurate for the contents that are changed with simple techniques. Cheng et al. [7] investigated by using text quality metrics if it is possible to identify antisocial users in their post history in online forums. A multi-label classifier trained by Gunasekara and Nejadgholi [8] for detecting toxicity in online conversational text, their result indicated that character-level text representation methods perform better than word-level representations. Hanu [9] developed a trained model to predict toxic contents named Detoxify. Detoxify method provides a toxicity score for each content to indicate whether the content is toxic or not and scores for different toxicity categories such as threats, obscenity, insults, and identity hate.

Prior works of some researchers indicate that they use different methods in various social media platforms to detect toxicity. DiCicco et al. [10] compared the toxicity between Parler and Twitter and analyzed the highly toxic users and their networks on these two platforms. Obadimu et al. [11] used an NMF method to predict commenter toxicity on YouTube. They claimed that the performance of the NMF model is more accurate than other models. Obadimu et al. [12], in their other study, focused on evaluating various forms of toxicity. They investigated their assumption on the YouTube comments posted on pro- and anti-NATO channels. In a similar study, Pascual-Ferrá et al. [13] evaluated the toxicity of Pro-mask and Anti-mask related to COVID-19 on Twitter. The finding indicated that Anti-mask hashtags are more toxic than Pro-mask.

Wallace Chipidza [14] discovered a network of content posted on 30 politically biased and two neutral subcommunities on Reddit. Related to COVID-19, his finding from graph modeling indicates that most highly toxic contents are likely to be in political subreddits. Rafal Urbaniak et al. [15] used algorithmic detection and Bayesian statistical methods, analyzed Reddit's contents to find the correlation between username toxicity and different types of that. On the other hand, Yun Yu Chong and Haewoon Kwak [16] discussed detecting toxicity triggers in an Asian online community and how they can differ from Western online communities. Hind Almerekhi et al. [17], in their study, investigated the detection of toxic contents and the source of the toxicity in the discussion on Reddit. For this aim, they propose an approach for toxic comment and toxicity trigger detection.

## III. DATA COLLECTION

The data from Twitter, Parler, and Reddit analyzed in this work consisting of a corpus of user posts collected based on a list of seed hashtags related to COVID-19 from January 1, 2020, through December 31, 2020 (Table 1).

A total of twelve datasets were created, four for each platform with mirroring hashtags and keywords. An open dataset from the Parler social network was created by Aliapoulios et al. [18], a complete dataset of all Parler data from August 2018 to when Parler was shut down in January 2021. The data for this paper was filtered by the seed list of keywords (Table 1). The Twitter data was collected using the Twitter Developer API [19] for the hashtags in (Table 1) posthoc. Because of this, tweets and accounts removed from Twitter for being labeled misinformation were not collected. Finally, Reddit posts and comments were collected using Pushshift API [20][21]. The customized python code was developed to collect data containing specific keywords during a specific period using the PSAW library [22]. Reddit data were collected from the whole of Reddit. A total of 72,327 posts and comments were collected from the 7511 subreddit. According to Twitter, Parler, and Reddit data-sharing guidelines, data collected in the study will be made available upon request.

TABLE 1. KEYWORDS USED FOR DATA COLLECTION WITH MEAN AND STANDARD DEVIATION OF TOXICITY SCORE.

| Categories | Social Media | Records | Mean Toxicity | SD |
|---|---|---|---|---|
| Covid | Twitter | 28,131 | 0.234 | 0.388 |
| | Parler | 16361 | 0.294 | 0.402 |
| | Reddit | 24501 | 0.1959 | 0.314 |
| Lockdown | Twitter | 1472 | 0.326 | 0.406 |
| | Parler | 5965 | 0.176 | 0.361 |
| | Reddit | 4781 | 0.216 | 0.32 |
| Mask | Twitter | 2423 | 0.313 | 0.416 |
| | Parler | 26165 | 0.264 | 0.388 |
| | Reddit | 16086 | 0.23 | 0.348 |
| Vaccine | Twitter | 610 | 0.302 | 0.411 |
| | Parler | 5928 | 0.119 | 0.304 |
| | Reddit | 26959 | 0.81 | 0.25 |

## IV. METHODOLOGY

Before executing the toxicity analysis, the seed keywords and hashtags from each record in the datasets were removed so their presence would not influence the calculated toxicity scores for the overall target corpus. After the toxicity analysis, non-English posts for Parler, Posts and comments for Reddit, tweets, and retweets for Twitter were removed as Detoxify Unified was only trying to support the English language. Because of this, the results in other languages could have been more accurate. There were some missing, deleted, removed, and duplicate posts and comments on Reddit. There were some duplicate posts and comments that contained multiple keywords that were being searched. So, every duplicate value was removed to ensure that all datasets contained the unique value. When the analysis was completed, we computed toxicity scores for each Parler post, Twitter tweet, and Reddit post and comment in the dataset using Detoxify. Detoxify, a model created by Unitary AI (https://github.com/unitaryai/detoxify), uses a Convolutional Neural Network. It is trained with word vector inputs to determine whether the text could be perceived as toxic to a discussion. Given a text input, the Detoxify API returns a probability score between 0 and 1, with higher values indicating a greater likelihood of the toxicity label being applied to the text. Since toxicity scores are based on a probability score of 0 to 1, toxicity scores of 0.5 or greater indicate a piece of text labeled as toxic. Detoxify returns seven categories of toxicity scores in terms of level and type 1) toxicity, which is the overall level of toxicity for a piece of text 2) severe toxicity 3) obscene 4) threat 5) insult 6) identity attack and 7) sexually explicit. Detoxify is used since it is an open-source comment detection python library that identifies harmful and inappropriate texts online. This multilingual model has been trained in English, French, Italian, Spanish, Russian, Turkish, and Portuguese. Even though it can predict toxicity by giving a score, it is not efficient, while some words related to swearing, insults, or profanity are present in the text. They may predict a non-toxic text as toxic if there are certain words. For comparison, we also explored using Google's Perspective API, a related model with similar outputs used for determining toxicity. Previous datasets for other research were analyzed using both tools to compute the toxicity scores, finding similar values for toxicity scores across the same dataset.

## V. ANALYSIS AND RESULTS

In this section, we present our analysis and results. First, we discuss the overall posting frequency of our seed hashtags (and keywords) and the results of our toxicity analysis for each platform, Twitter, Parler, and Reddit.

For the Twitter dataset, the seed hashtags used in this analysis first appeared in March 2020. Of all the Twitter datasets, COVID had the most posts from March through December 2020. There was a peak in mid-April and near the end of June, and then a significant rise in the number of tweets in mid-November.

For Parler, the seed keywords (to mirror the Twitter target hashtags) registered posting activity near the end of May. Interestingly, all Parler datasets simultaneously registered a huge spike that peaked and then fell in posting frequency during November. This is a curious result that may indicate inorganic behavior at first glance. Further inspection of the dataset revealed that Parler users often adopted the behavior of using all four seed hashtags within a single post, which was not the behavior of Twitter users.

For Reddit, the number of posts and comments started to show up in the early weeks of 2020, which is earlier than Twitter and Parler. This is because some subreddit named 'r/worldnews' and 'r/China_Flu' have started discussions about COVID-19 since it first spread in China in Late December March. Each keyword-related post peaked from late November to early December.

Although each keyword or hashtag containing posts, comments, and tweets follow almost the same weekly trend throughout the year, three different platforms have different trends for different keywords. For instance, Twitter datasets had more tweets related to the f*ckcovid hashtag, whereas Parler had more posts containing the keyword f*ckmask. On the other hand, if we consider Reddit posts and comments, the f*ckvaccine keyword containing posts and comments was in the lead.

Thus, Twitter is more toxic based on COVID-related tweets, and Parler is more toxic for mask-related posts. Finally, Reddits' toxicity is mostly based on vaccine-related posts and comments. Multiple subreddit like 'r/Nonewnormal' got banned due to spreading misinformation about vaccination during that time. We have collected posts and comments from that subreddit if they contain those four keywords related to COVID-19. Even though some subreddit got banned due to violation of community guidelines on Reddit. The posts and comments are collected using Pushshift API and analyzed toxicity on those posts.

As mentioned above, before executing toxicity analysis, these seed hashtags (mirroring keywords) were removed from each data record to avoid influencing the calculated toxicity scores for the overall target corpus. Upon completing our toxicity analysis methodology, we discovered that Twitter, Parler and Reddit differed in the existence of toxicity within their respective user-generated text content (toxicity scores > 0.5) from January 1, 2020, through December 31, 2020. When breaking down the content containing toxicity on each platform, Reddit contained a higher overall percentage, around 37% for all datasets, compared to Twitter, with just above 30%, and Parler, with 21.83%. (Table 2).

Although Reddit has the highest percentage of toxic posts (Toxicity score > 0.5), Twitter has the highest number of toxic posts containing the f*cklockdown hashtag, with 34.31% of tweets. In addition, Parler has 30.51% of the toxic post containing the keyword f*ckcovid.

However, surprisingly Reddit has 86% of toxic posts and comments containing the keyword f*ckvacccine, which is the highest among all platforms and all other hashtags and keywords. This made Reddit more toxic than the other two platforms. It is because Reddit has some forums that talk most about anti-vaccine.

TABLE 2. NUMBER AND PERCENTAGE OF TOXIC POSTS ON TWITTER, PARLER AND REDDIT FOR ALL TWELVE DATASET.

| Dataset | Platform | Total Tweets/Posts | Percentage of Post with Toxicity Score > 0.5 | | | Percentage of Posts with Toxicity score > 0.7 | | | Percentage of Posts with Toxicity Score > 0.9 | | |
|---|---|---|---|---|---|---|---|---|---|---|---|
| | | | Toxicity | Obscene | Insult | Toxicity | Obscene | Insult | Toxicity | Obscene | Insult |
| #f*ckcovid | Twitter | 28131 | 24.08% | 22.05% | 10.93% | 21.65% | 19.89% | 8.60% | 17.00% | 10.06% | 6.93% |
| #f*cklockdown | Twitter | 1472 | 34.31% | 28.60% | 16.37% | 27.45% | 22.96% | 11.35% | 20.11% | 14.54% | 6.05% |
| #f*ckmask | Twitter | 2423 | 31.24% | 23.15% | 19.81% | 27.90% | 20.59% | 16.05% | 22.86% | 15.44% | 5.94% |
| #f*ckvaccine | Twitter | 610 | 30.98% | 23.28% | 19.51% | 27.21% | 19.84% | 14.75% | 20.98% | 14.26% | 7.21% |
| #f*ckcovid | Parler | 16361 | 30.51% | 20.61% | 18.01% | 29.08% | 19.16% | 15.82% | 15.48% | 10.43% | 7.03% |
| #f*cklockdown | Parler | 5956 | 18.11% | 13.06% | 12.14% | 17.45% | 12.98% | 11.90% | 13.73% | 8.14% | 8.04% |
| #f*ckmask | Parler | 26165 | 26.80% | 15.71% | 17.12% | 23.38% | 13.28% | 13.48% | 14.64% | 9.96% | 7.18% |
| #f*ckvaccine | Parler | 5928 | 11.93% | 9.06% | 5.36% | 10.90% | 8.92% | 4.82% | 10.37% | 8.52% | 4.28% |
| #f*ckcovid | Reddit | 24501 | 18.41% | 13.14% | 6.53% | 13.35% | 8.63% | 4.11% | 7% | 3.64% | 2.11% |
| #f*cklockdown | Reddit | 4781 | 20.08% | 13.77% | 6.60% | 13.99% | 9.10% | 3.88% | 7.12% | 3.34% | 1.86% |
| #f*ckmask | Reddit | 16086 | 23.37% | 16% | 10.64% | 18.17% | 11.29% | 7.60% | 10.52% | 4.97% | 4.28% |
| #f*ckvaccine | Reddit | 26959 | 86% | 81.50% | 39.99% | 77.67% | 70.33% | 29.94% | 57.27% | 44.42% | 18.37% |

There was a huge community that discussed the covid vaccine. These subreddits are responsible for spreading misinformation related to the Covid vaccine. Though this subreddit eventually got banned, we collected posts from those banned subreddit. Most of the Twitter content had a higher probability of being labeled as toxic than the Parler and Reddit content, except the f*ckvaccine keyword for Reddit. Surprisingly, for the overall toxicity category, the Twitter content for all datasets had a higher percentage of content with toxicity scores greater than 0.7 and greater than 0.9 than did the Parler content and Reddit content. Again, Parler only exceeded Twitter in the percentage of harmful content for the COVID dataset. In contrast, Reddit exceeded the other two platforms in the percentage of harmful content for the vaccine dataset.

This is an interesting result because we expected to see more harmful content on Parler due to the free-speech nature of the platform and how they tout their lack of censorship as a selling point for users. We also expected to see the highest toxicity on Reddit for the vaccine dataset. We also looked at the obscene and insult toxicity categories for each tweet and post for all twelve datasets. Of the seven categories of toxicity scores obtained from Detoxify, only three contained enough data to warrant inclusion in the discussion: toxicity (overall), obscene, and insult. More Twitter content fell into the obscene category than did Parler and Reddit content for all datasets except the vaccine dataset from Reddit, with the highest percentage being within the Lockdown dataset (28.6% for Twitter, 13.06% for Parler and 13.77% for Reddit) and vaccine dataset (23.28% for Twitter 9.06% for Parler and 81.50% for Reddit). However, more Parler content fell into the insult category than Twitter content and Reddit content for the COVID dataset (18.01% vs. 10.93% vs. 6.53%).

The percentage of harmful content (overall toxicity category) within the vaccine datasets varied considerably between platforms (30.98% for Twitter versus 11.93% for Parler versus 86% for Reddit). So, overall, the toxicity analysis revealed that Twitter was more toxic than Parler and Reddit in all, but one case, the COVID dataset and Reddit were more toxic than Parler and Twitter for the vaccine dataset. The toxic content was more obscene and insult type for both platforms. However, the harmful content on Twitter was obscener than that of Parler, especially within the Lockdown dataset. The toxic content on Parler was more of an insulting type within the COVID dataset. Finally, the vaccine dataset on Reddit contained the highest toxic, obscene, and insulting posts than the other two platforms.

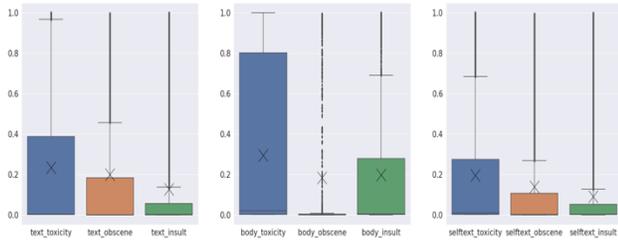

Figure 1. f*ckcovid hashtag for three classes (Toxicity, Obscene, Insult) for Twitter (left) vs Parler (middle) vs Reddit (right).

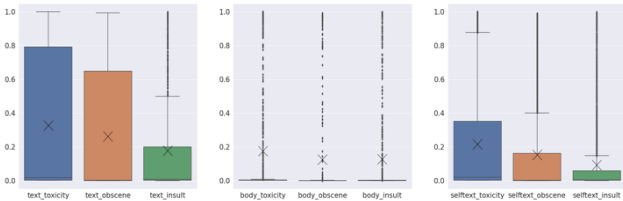

Figure 2. f*cklockdown hashtag for three classes (Toxicity, Obscene, Insult) for Twitter (left) vs Parler (middle) vs Reddit (right).

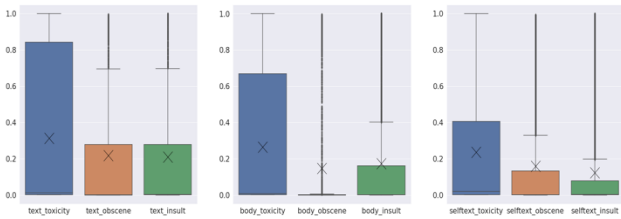

Figure 3. f*ckmask hashtag for three types of classes (Toxicity, Obscene, Insult) for Twitter (left) vs Parler (middle) vs Reddit (right).

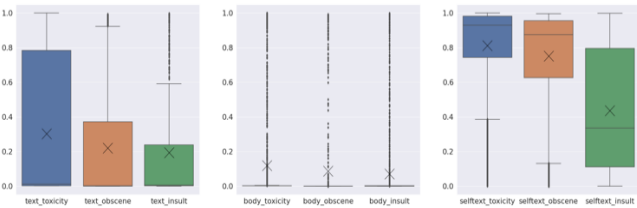

Figure 4. f*ckvaccine hashtag for three types of toxicity (Toxicity, Obscene, Insult) for Twitter (left) vs Parler (middle) vs Reddit (right).

The Twitter data, for example, shows that a few conversations are very toxic, and those few highly toxic conversations are driving up the overall toxicity level of the platform. The same goes for the Reddit vaccine dataset as well. This has important implications for platform administrators, who can significantly reduce the strongest drivers of toxicity by moderating the relatively few, highly toxic users rather than attempting larger platform-wide changes to all users. The toxicity standard deviation metrics revealed some unique contrasts between the platforms (Table 1). The standard deviation of toxicity values for content within the lockdown, mask, and vaccine categories are higher on Twitter than on Parler and Reddit, indicating that there is more variation in toxicity for these datasets. However, values were higher for Parler for content within the COVID category.

Figure 1 to Figure 4 illustrate that the term f*ckcovid on Parler is generally more toxic than on Twitter and Reddit. The mean toxicity is marked by a cross on each boxplot, slightly higher than Parler. However, for the other terms except for vaccine, Twitter is more toxic. All five points for the vaccine dataset on Reddit are the highest among all platforms. From the seven toxicity classes, we take three severe classes to compare in our statistical analysis. For F*Lockdown hashtags, Parler and Reddit are less toxic than Twitter if we consider the mean toxicity from the boxplot for both platforms. On the other hand, for f*ckcovid and f*ckmask hashtags, there is a significant increase in toxicity in Parler. On Twitter, the most toxic term f*ckmask whereas for Parler, it is f*ckcovid, and on Reddit, it is f*ckvaccine.

## VI. CONCLUSIONS AND FUTURE WORKS

Twitter and Parler both experienced moderate levels of toxicity regarding COVID-19 content. However, Reddit had the highest toxicity related to the vaccine dataset, which is much higher than any other keywords or platforms. This paper compares and analyzes the toxicity of these three social media platforms in the same period. The methods were applied to different datasets for Twitter, Parler, and Reddit. The key finding of this research indicates Reddit is the most toxic social media platform among these three and Parler contained less toxicity compared to Twitter and Reddit regarding COVID-19 discourse.

Although the finding indicates toxicity levels were higher overall on Twitter for all datasets except for COVID-19 and vaccine, it was surprising to observe higher toxicity levels on Twitter since it is a moderated platform with clear guidelines for content posted, whereas Parler's guidelines emphasize a lack of moderation. Even though Reddit experienced the highest toxicity for the vaccine topic, the moderators took necessary steps to decrease toxicity by banning the anti-vaccine subreddit named 'r/Nonewnormal'. One possible explanation for the unexpectedly high toxicity on the Parler COVID dataset is that Twitter began removing users and posts sharing COVID-19 misinformation in April 2020, sparking anger and prompting many users to migrate to Parler instead [23]. In addition to being detrimental to the overall health of social networks, the moderate proportion of harmful content on these platforms surrounding COVID-19 topics may have affected users' perceptions of the effectiveness and importance of periodic lockdowns, wearing of face masks, and becoming vaccinated. The contributions of this work include evidence that 1) Twitter contained a higher level of toxicity regarding COVID-19 discourse than did Parler and Reddit; 2) Reddit contained the highest level of toxicity among all three social platforms for vaccine-related discussion. 3) Parler contained the highest level of toxicity among all three social platforms for COVID-related discussion.

A potential limitation of this paper is the methodology used to collect and analyze the data - the seed hashtag stem #f*ck can be used positively or negatively, depending on the context. The model used in this paper to classify content as toxic or not has difficulty distinguishing the semantic context of profanity and often classifies profane words as toxic, regardless of intent. We will keep this limitation in mind going forward in our future works. In future work, we plan to expand our keywords and collect more data from these three platforms, which are easy to get under their guidelines. We are working on other popular social media platforms like TikTok and Facebook. In addition, we will further explore the vaccine and lockdown topics due to their notably higher toxicity on Reddit.

ACKNOWLEDGEMENT

This research is funded in part by the U.S. National Science Foundation (OIA-1946391, OIA-1920920, IIS-1636933, ACI-1429160, and IIS-1110868), U.S. Office of the Under Secretary of Defense for Research and Engineering (FA9550-22-1-0332), U.S. Office of Naval Research (N00014-10-1-0091, N00014-14-1-0489, N00014-15-P-1187, N00014-16-1-2016, N00014-16-1-2412, N00014-17-1-2675, N00014-17-1-2605, N68335-19-C-0359, N00014-19-1-2336, N68335-20-C-0540, N00014-21-1-2121, N00014-21-1-2765, N00014-22-1-2318), U.S. Air Force Research Laboratory, U.S. Army Research Office (W911NF-20-1-0262, W911NF-16-1-0189, W911NF-23-1-0011), U.S. Defense Advanced Research Projects Agency (W31P4Q-17-C-0059), Arkansas Research Alliance, the Jerry L. Maulden/Entergy Endowment at the University of Arkansas at Little Rock, and the Australian Department of Defense Strategic Policy Grants Program (SPGP) (award number: 2020-106-094). Any opinions, findings, and conclusions or recommendations expressed in this material are those of the authors and do not necessarily reflect the views of the funding organizations. The researchers gratefully acknowledge the support.